\RequirePackage{fix-cm}
\documentclass[twocolumn,epjc3]{svjour3}
\smartqed  

\RequirePackage{amsmath,amssymb,amsfonts}
\RequirePackage{color}

\RequirePackage{graphicx}
\journalname{Eur. Phys. J. C}
\begin{document}

\title{Parton shower evolution in medium and nuclear modification of photon-tagged jets in Pb+Pb collisions at the LHC
}


\author{Guang-You Qin 
}



\institute{Institute of Particle Physics, Central China Normal University, Wuhan, 430079, China 
           \and
 	Department of Physics and Astronomy, Wayne State University, Detroit 48201, USA 
           \and
	Department of Physics, Duke University, Durham, North Carolina 27708, USA
}

\date{Received: date / Accepted: date}

\maketitle

\begin{abstract}

We study the medium modification of jets correlated with large transverse momentum photons at the LHC via a transport and perturbative QCD hybrid model which incorporates the contributions from both elastic collisions and radiative energy loss experienced by the parton showers.
Calculations are performed for the modification of the photon-tagged jet yield, the photon-jet energy imbalance, and the azimuthal distribution of away-side jets.
The modifications of photon-tagged jets with different $x_T = p_{T, J}/p_{T, \gamma}$ values are studied and they exhibit different centrality and jet cone size dependence due to traversing different medium lengths and density profiles.
We further investigate the influence of transverse and longitudinal jet transport coefficients on the nuclear modification of photon-tagged jet production and jet shape observables.

\end{abstract}

\section{Introduction}
\label{intro}

The energetic probes produced in early hard scattering processes provide valuable tools for studying the properties of the quark-gluon plasma (QGP) created in relativistic heavy-ion collisions.
Large transverse momentum ($p_T$) jets are particularly interesting as they directly interact with the constituents of dense QCD matter, and are modified in the process \cite{Majumder:2010qh}.
Jet quenching has been confirmed by multiple experimental observations, such as the suppression of single inclusive hadron  production at high $p_T$ in nucleus-nucleus collisions compared to binary collision scaled proton-proton collisions at the same energies \cite{Adcox:2001jp,Adler:2002xw,Aamodt:2010jd}.

A lot of theoretical effort has been devoted to understand the modification of jets in hot and dense strong-interaction QCD matter.
The medium modification of jets originates from a combination of elastic collisions with the medium constituents and induced gluon emission while jets propagate through such highly excited nuclear medium \cite{Bjorken:1982tu,Braaten:1991we,Zakharov:1996fv,Baier:1996kr}.
There are currently a few perturbative QCD based schemes for calculating radiative and collisional parton energy loss \cite{Gyulassy:1999zd,Wiedemann:2000za,Arnold:2001ba,Wang:2001ifa,Majumder:2009ge,Bjorken:1982tu,Braaten:1991we,Mustafa:2004dr,Peshier:2006hi,Qin:2007rn}.
A systematic comparison of these different jet quenching schemes has been carried out in Ref. \cite{Armesto:2011ht}  in the context of the ``brick" problem.
Various phenomenological studies have been performed for the suppression of single inclusive high $p_T$ hadron production \cite{Bass:2008rv,Wicks:2005gt,Qin:2007rn,Armesto:2009zi,Chen:2010te}, nuclear modification of dihadrons \cite{Zhang:2007ja,Renk:2008xq}, photon-hadron \cite{Renk:2006qg,Qin:2009bk,Zhang:2009rn} correlations in high energy nucleus-nucleus collisions, etc.

In recent years, sophisticated experimental techniques have been utilized to reconstruct full jets emitted in relativistic heavy-ion collisions, providing both challenges and opportunities for our understanding of jet modification in dense nuclear matter in terms of both leading and subleading fragments of the jet showers \cite{Vitev:2009rd}.
A few Monte-Carlo generators have also been developed to simulate the modification of full jet propagation through dense nuclear medium \cite{Armesto:2009fj,Zapp:2009ud,Schenke:2009gb,Renk:2010zx,ColemanSmith:2011rw,Majumder:2013re,Wang:2013cia}.

Meanwhile, the large kinematics available at the LHC allows us to investigate medium effects on jets with transverse energies over a hundred GeV.
The strong modification of the momentum imbalance
distribution between the correlated jet pairs first measured in Pb+Pb collisions at the LHC \cite{Aad:2010bu,Chatrchyan:2011sx} indicated that the away-side subleading jets experienced significant energy loss during their propagation through the produced hot and dense medium.
These results have triggered wide interest in studying full jets and their in-medium evolution and modification as opposed to single particle observables.
Many calculations have suggested that significant jet cone energy loss and the observed dijet asymmetry may come from the loss of low energy gluons from the jet cone to the medium and are then not recovered by the jet algorithm. These gluons escape either by deflection or by energy dissipation and thermalization due to the interaction with the medium constituents  \cite{CasalderreySolana:2010eh,Qin:2010mn,Lokhtin:2011qq,Young:2011qx,He:2011pd,Renk:2012cx,CasalderreySolana:2012ef,Blaizot:2013hx}.

Another exciting result is the measurement of full jets correlated with high $p_T$ photons in Pb+Pb collisions at the LHC \cite{Chatrchyan:2012gt}. Jets tagged by photons and electro-weak bosons have been regarded as the ``golden" channel for studying jet quenching due to the fact that the boson triggers, once produced, will escape the medium without further interaction.
Thus they are expected to put a stronger constraint on the momentum distributions of the away-side tagged jets than hadron or jet triggers \cite{Wang:1996yh}.
Strong modification of the photon-jet momentum imbalance distribution has been obtained from model calculations as well \cite{Neufeld:2012df,Dai:2012am}.

The goal of various jet quenching studies is to acquire a better understanding of jet-medium interaction and obtain the transport properties of the hot and dense matter created in relativistic nuclear collisions.
Much of current attention has been focused on the quantitative extraction of jet transport coefficients, such as $\hat{e}=dE/dt$, $\hat{e}_2 = d(\Delta E)^2/dt$ and $\hat{q} = d(\Delta p_T)^2/dt$ \cite{Baier:1996kr,Majumder:2008zg,Qin:2012fua}.
These coefficients not only control how energy and momentum are lost from jets and dumped into the medium \cite{Qin:2009uh,Neufeld:2009ep,Li:2010ts}, but are necessary ingredients for the investigation of medium response to jet propagation.
The knowledge of these coefficients and their (relative) sizes may provide much insight into the internal structures and the properties of the hot and dense matter probed by the high energy jets.

With this motivation, we develop a Monte-Carlo transport and perturbative QCD hybrid model to simulate jet shower evolution and jet modification in dense QCD medium and apply it to study the medium modification of jets correlated with high $p_T$ photons at the LHC energies.
The effect of elastic collisions on the parton shower is encoded by the above mentioned transport coefficients, namely the longitudinal drag and diffusion as well as transverse momentum broadening.
The same transport coefficients also control the radiative energy loss which is simulated by employing the single gluon emission spectrum obtained from the perturbative QCD-based higher twist jet energy loss calculation \cite{Wang:2001ifa,Majumder:2009ge}.

Our paper is organized as follows. In section II, we provide some details about how we simulate the evolution and energy loss of jet showers in medium. The contributions of elastic collisions and gluon radiation to jet energy loss are compared.
The numerical calculations for photon-jet correlations in Pb+Pb collisions at the LHC energies are presented in section III, including the nuclear modification of tagged jet yield, the photon-jet momentum imbalance and the azimuthal distribution of the away-side jets.
To study the medium effects on jets traversing different medium lengths and density profiles, we compare the nuclear modification of photon-tagged jets with different values of $x_T = p_{T, J}/p_{T, \gamma}$.
We further investigate the influence of longitudinal and transverse jet transport coefficients on the nuclear modification of photon-tagged jets.
The last section contains our summary.

\section{Parton shower evolution in medium}

In this section, we describe our Monte-Carlo method for simulating the transport and modification of a partonic jet shower in dense QCD matter.
We assume each parton of the jet shower follows classical trajectories between each jet-medium interaction that induces momentum change of the parton (either by an elastic collision or the radiation of a gluon).
We keep track of all the propagating partons and the radiative gluons until they exit the dense medium.

The probability for a parton at each time step $\Delta t$ to have an momentum exchange by an elastic collision with the medium constituents is determined by,
\begin{eqnarray}
P_{\rm coll}(t, \Delta t) = \Delta t / t_{\rm mf},
\end{eqnarray}
where $t_{\rm mf}$ is an ``effective" mean free time, thus $1/t_{\rm mf}$ is the rate of the elastic collisions (the exchange of energy and momentum with medium).
If there is a jet-medium interaction through an elastic collision, the energy and momentum exchange is sampled from a probability distribution of $P(\Delta E, \Delta \vec{p}_\perp, t, \Delta t)$.

Generally, the above mean free time $t_{\rm mf}$ and the distribution of the momentum and energy exchange can be determined if the details about the medium structure and jet-medium interaction are known.
For example, the mean free time can be obtained from the density $\rho$ of the medium and the cross section $\sigma$, i.e., $t_{\rm mf} = 1/(\rho \sigma)$. In this work, we take the multiple soft scattering limit in which the momentum exchange through each elastic collision is small.
In this limit, the probability distribution $P(\Delta E, \Delta \vec{p}_\perp, t, \Delta t)$ is a Gaussian for both longitudinal and transverse directions.
Thus we may set the ``effective" mean free time in our simulation equal to the time step ($t_{\rm mf}=\Delta t$), i.e., in each time step, the parton will experience a momentum exchange due to a soft elastic scattering.
The mean and variance of the distribution are then determined by the longitudinal drag $\hat{e} t_{\rm mf}$, longitudinal diffusion $\hat{e}_2 t_{\rm mf}$ (and transverse broadening $\hat{q} t_{\rm mf}$).
We have checked that the variation of $t_{\rm mf}$ values does not affect the results if a Gaussian form is used for the momentum exchange in a sufficiently long medium ($\Delta t \leq t_{\rm mf} \ll L$).
We note that within our framework one may go beyond the multiple scattering limit, such as a few hard scatterings, with the use of more realistic values of $t_{\rm mf}$ and non-Gaussian form of momentum exchange \cite{Schenke:2009ik,Auvinen:2011aa}.
We leave this to a future study.

We simulate the radiation of gluons by employing the single gluon emission spectrum from higher twist calculations \cite{Wang:2001ifa,Majumder:2009ge},
\begin{eqnarray}
\frac{dN_g}{dx dl_\perp^2 dt} = \frac{2\alpha_s}{\pi} P(x) \frac{\hat{q}_A}{l_\perp^4} \sin^2\left(\frac{t-t_i}{2t_{\rm form}}\right),
\label{dNg_dxdlt2dt}
\end{eqnarray}
where $x$ and $l_\perp$ are the fractional energy and transverse momentum carried by the radiated gluon, $P(x)$ is the vacuum parton splitting function, $\hat{q}_A$ is the gluon transport coefficient, and $t_{\rm form}=2Ex(1-x)/l_\perp^2$ is the gluon formation time.
In the above equation, the gluon radiation is purely stimulated by the transverse momentum broadening of the jet.
Generally, the longitudinal momentum exchange (drag and diffusion) may give rise to medium-induced gluon radiation as well.
Such contribution is straightforward to include in the current Monte-Carlo simulation when it becomes available \cite{Qin:}.

The probability for a parton in each time step $\Delta t$ to radiate a gluon is determined from medium-induced gluon radiation spectrum:
\begin{eqnarray}
P_{\rm rad}(t, \Delta t) = 
\langle N_g (t, \Delta t)\rangle = \Delta t \int dx dl_\perp^2 \frac{dN_g}{dx dl_\perp^2 dt}. \ \ \
\end{eqnarray}
We choose the value of $\Delta t$ to ensure that the average radiated gluon number is smaller than 1 in a time step $\Delta t$ (0.01~fm in our simulation).
For $\langle N_g (t, \Delta t)\rangle \ll 1$, our treatment is equivalent to the Poisson distribution for the gluon emission ($1-e^{- \langle N_g \rangle} \approx \langle N_g \rangle)$.

If there is a gluon emitted in a given time step, the energy and the momentum of the radiated gluon are sampled from the following probability distribution,
\begin{eqnarray}
P(x, l_\perp^2, t, \Delta t) = \frac{\Delta t}{\langle N_g \rangle} \frac{dN_g}{dx dl_\perp^2 dt}.
\end{eqnarray}
When simulating jet shower propagating through the hot and dense medium, we allow the radiated gluons to interact with the medium after their formation times $t_{\rm form}$ are reached.
We impose a minimum energy cutoff $\pi T$ for the medium-induced radiation spectrum in order to take into account the balance between gluon radiation and absorption from the hydrodynamic medium.
To simulate multiple gluon emission, the initial time $t_i$ in the above equation is reset to zero when a gluon is radiated so that the probability of radiating another gluon starts to accumulate again with time.
We note that the framework described here may be applied to other energy loss formalisms as well, as long as the energy/momentum distribution of the radiated gluons is provided.

In our simulation, the evolution of parton showers in dense QGP medium are controlled by three transport coefficients $\hat{e}$, $\hat{e}_2$ and $\hat{q}$.
These coefficients can be determined if detailed information about medium structure and jet-medium interaction are known \cite{Majumder:2012sh}.
In this work, the longitudinal drag and diffusion are related by applying the fluctuation-dissipation relation $\hat{e}_2 \approx 2T\hat{e}$.
Unless otherwise stated, we assume the diffusion in transverse and longitudinal directions to be equal, i.e, $\hat{q} \approx 2 \hat{e}_2$,
We will investigate the influence of varying relative sizes of transverse and longitudinal transport coefficients on the jet modification in the last section.

\begin{figure}[tb]
\includegraphics[width=0.95\linewidth]{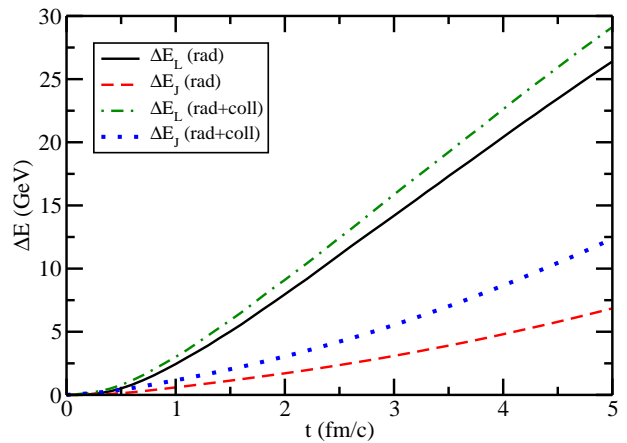}
 \caption{(Color online) Comparison of full jet ($R=0.3$) energy loss and leading parton energy loss as a function of time with/without the inclusion of the contribution from elastic collisions.
} \label{dE_vst}
\end{figure}

In Fig. \ref{dE_vst}, we show the energy loss experienced by a quark jet propagating through a static quark gluon plasma with a temperature of $400$~MeV.
A quark with initial energy of $100$~GeV enters into the medium at $t=0$, and a shower of partons will develop due to the medium-induced gluon radiation.
In this plot, the jet transport coefficient is set as $\hat{q} = 2\hat{e}_2 = 4T\hat{e} = 2$~GeV/fm$^2$.
We show the energy loss experienced by the leading parton and the full jet shower (with cone size $R=0.3$), with and without the inclusion of the contribution from elastic collisions.

The energy loss of the leading quark is found to be dominated by radiative energy loss; the inclusion of collisional energy loss does not have much effect on leading parton energy loss.
However, the energy loss of the full jet is increased by almost a factor of 2 when the contribution from elastic collisions is taken into account. This is due to the fact that for full jet shower energy loss, one needs to take into account not only the collisional energy loss from the leading parton, but the effect of elastic collisions on radiated partons as well. Such an effect accumulates with time as a result of more and more partons in the shower being developed by the medium-induced radiation. This leads to stronger length dependence for full jet energy loss than leading parton energy loss. Similar effect has been found in Ref. \cite{Qin:2009uh,Neufeld:2009ep}.

\section{Nuclear modification of photon-tagged jets}

In this section, we present the numerical results for the medium modification of jets correlated with high $p_T$ photons in Pb+Pb collisions at $\sqrt{s_{NN}} = 2.76 {\rm TeV}$ at the LHC.
To simulate the medium modification of jet shower in relativistic heavy-ion collisions, we need to provide the initial conditions before the jet-medium interaction.
In this application, the production of the photon-jet pairs from early hard scattering processes are generated from PYTHIA8.1 \cite{Sjostrand:2007gs} for p+p collisions at $\sqrt{s_{NN}} = 2.76 {\rm TeV}$.
We use anti-$k_T$ algorithm to reconstruct the full jets via the FASTJET package \cite{Cacciari:2011ma}, from which we also obtain the detailed content of the full jets, such as the energy and momentum for each parton in a jet.
The reconstructed vacuum jets are provided as inputs to our Monte-Carlo code which simulates in-medium jet evolution and medium modification.
We note that during the propagation of jets through medium, energy can flow in and out of jet cone.
To take into account such effect, when studying the medium effect for jets with given size (e.g., $R=0.3$), the vacuum jets with a larger jet cone size (e.g., $R=1$) are used as initial conditions for simulating medium modification.

The initial photon-jet production points are sampled according to the distribution of the binary collisions obtained from a Glauber model simulation.
The space-time evolution profiles (energy/entropy density, temperature and flow velocities) of the bulk QGP medium that jets interact with are provided by relativistic hydrodynamics simulation.
In this work, we employ a (2+1)D viscous hydrodynamics model (VISH2+1) developed by The Ohio State University group \cite{Song:2007fn,Song:2007ux,Qiu:2011hf}, with two-component Glauber model for hydrodynamics initial conditions.
The code version and parameter tunings for Pb+Pb collisions at LHC energies are taken as in Ref. \cite{Qiu:2011hf}.
When the local temperature of the medium drops below the transition temperature of $160 {\rm MeV}$, jets are decoupled from the medium.
After jets exit the hydrodynamic medium, detailed information of the modified jets is again provided to the FASTJET package to perform the full jet reconstruction (to be compared with these jets without medium modification).
In this work, we do not perform the hadronization process for the parton showers which we postpone to a future effort.

To study the medium modification on the photon-tagged jets, we may define two distribution functions which differ only by normalization,
   \begin{eqnarray}
   f(x_T) &=& ({1}/{N_\gamma}) {dN_{J\gamma}}/{dx_T}, \nonumber \\
	    P(x_T) &=& ({1}/{N_{J\gamma}}) {dN_{J\gamma}}/{dx_T},
   \end{eqnarray}
   where $x_T$ is the tagged jet momentum fraction $x_T = p_{T,J} / p_{T, \gamma}$.
   Given the kinematic cuts for the trigger photons and associated jets, the former distribution is normalized to the fraction of photons having the associated jet pair $R_{J\gamma}$ and the latter gives unity when integrating out the momentum fraction $x_T$.
   The nuclear modification factor $I_{AA}$ for the tagged jet distribution is usually defined as
   \begin{eqnarray}
   I_{AA}(x_T) &=& f_{AA}(x_T)/f_{pp}(x_T).
   \end{eqnarray}

   \begin{figure}[tb]
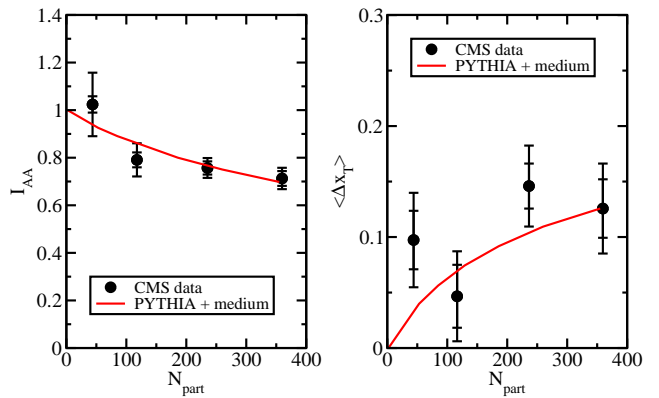

   \includegraphics[width=0.49\linewidth]{2a_gammajet_Iaa_vsnpart.eps}
   \includegraphics[width=0.49\linewidth]{2b_gammajet_dxT_vsnpart.eps}
   \caption{(Color online) The nuclear modification factor $I_{AA}$ and the average fractional momentum loss $\langle \Delta x_T \rangle$ for photon-triggered jets as a function of centrality for Pb+Pb collisions at the LHC. The jet size is $R=0.3$.
   } \label{gammajet_IaadxT}
\end{figure}

In Fig. \ref{gammajet_IaadxT} we show the $x_T$-integrated nuclear modification factor $I_{AA}=R_{J\gamma}^{AA}/R_{J\gamma}^{pp}$ (left) and the average fractional momentum loss $\langle \Delta x_T \rangle = \langle x_T \rangle|_{pp} - \langle x_T \rangle|_{AA}$ (right) as a function of centrality (the participant number $N_{\rm part}$).
The data points shown here are obtained from CMS measurements of $R_{J\gamma}$ and $\langle x_T \rangle$ for Pb+Pb collisions and PYTHIA+HYDJET references \cite{Chatrchyan:2012gt}.
For all the presented results in this work, we take the CMS kinematic cuts, i.e., the photon momentum $p_{T, \gamma} > 60 {\rm GeV}$, jet momentum $p_{T, J} > 30 {\rm GeV}$, and the azimuthal angle between trigger photons and associated jets $\Delta \phi = |\phi_\gamma - \phi_J| < 7\pi/8$.

Here we scale the transport coefficients to the medium local temperature according to their dimensions $\hat{q} = 2\hat{e}_2 = 4T\hat{e} \propto T^3$, with the overall normalization to be fitted to one data point.
In next section, we will investigate the influence of relative sizes and different parametrization forms of jet transport coefficients on medium modification of photon-tagged jets.
By fitting to the value of $I_{AA} \approx 0.7$ in most central ($0-10\%$) Pb+Pb collisions; we obtain the gluon transport coefficient $\hat{q}_{A} = 6.5 {\rm GeV}^2/{\rm fm}$ at the hydrodynamics initial time $\tau_0=0.6 {\rm fm}/c$ at the LHC (which corresponds to gluon $\hat{q}_{A} = 2.9 {\rm GeV}^2/{\rm fm}$ at the same initial time at RHIC).
Comparable values of jet transport coefficients have been obtained from other model calculations \cite{Bass:2008rv,Majumder:2011uk}.
Due to stronger medium effect from peripheral to central collisions, we observe the decrease of the nuclear modification factor $I_{AA}$ and the increase of the average fractional momentum loss $\langle \Delta x_T \rangle$ for photon-triggered jets.

\begin{figure}[tb]
\includegraphics[width=0.95\linewidth]{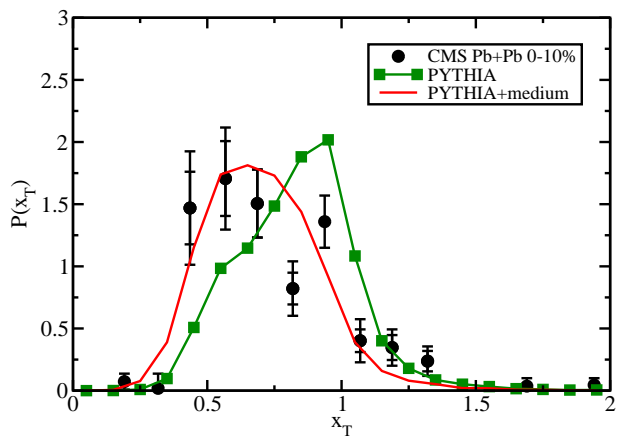}
\caption{(Color online) The distribution of the momentum imbalance variable $x_T$ between triggered photons and associated jets for most central ($0-10\%$) Pb+Pb collisions at the LHC. The jet size is $R=0.3$.
} \label{gammajet_xT}
\end{figure}

\begin{figure}[tb]
\includegraphics[width=0.95\linewidth]{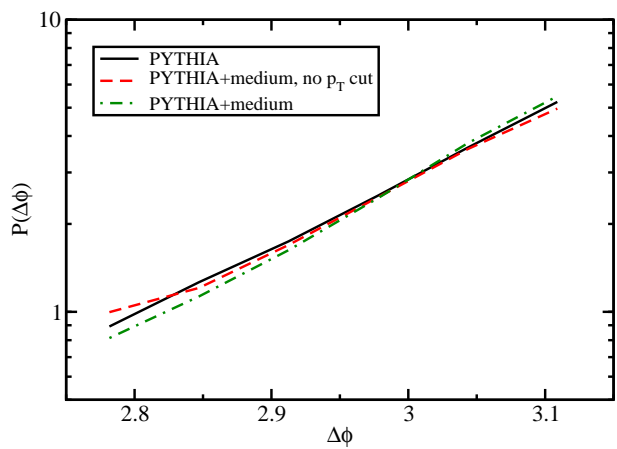}
\caption{(Color online) The distribution of the azimuthal angle $\Delta \phi$ between triggered photons and associated jets for most central ($0-10\%$) Pb+Pb collisions at the LHC. The jet size is $R=0.3$.
} \label{gammajet_dphicut}
\end{figure}

The distribution $P(x_T)$ of the momentum imbalance variable $x_J$ between triggered photons and the away-side jets is shown in Fig. \ref{gammajet_xT}. One observes a shift of the distribution to smaller $x_J$ values for the distribution in Pb+Pb collisions compared to p+p collisions. This indicates that the away-side tagged jets lose some amount of energy during their propagation through the hot and dense QGP created in the Pb+Pb collisions.

The azimuthal angle between trigger photons and associated jets $\Delta \phi = |\phi_\gamma - \phi_J|$ is another interesting quantity. In Fig. \ref{gammajet_dphicut}, the normalized distribution $P(\Delta \phi)$ is shown for both p+p collisions and most central ($0-10\%$) Pb+Pb collisions.
While there is strong medium effect for the momentum imbalance $x_J$ distribution as shown in Fig. \ref{gammajet_xT}, no strong modification is observed for photon-jet azimuthal angle $\Delta \phi$ distribution.
This is partly due to the kinematic cuts applied on the selection of photon-jet events.
Strongly deflected jets are usually those having experienced larger energy loss in the medium, and many of them may not be selected in the final presented results after applying the kinematic cuts.
Such effect may be seen from the dashed curve which represents the results for events with no $p_T$ cut applied to the associated jet momentum after the in-medium evolution.
One also observes some broadening for photon-jet azimuthal angle $\Delta \phi$ distribution for this case.

\begin{figure}[tb]

\setlength{\unitlength}{1mm}
\begin{picture}(90,90)

	\put(42,-1){x(fm)}

	\put(2,46){\includegraphics[width=4.2cm]{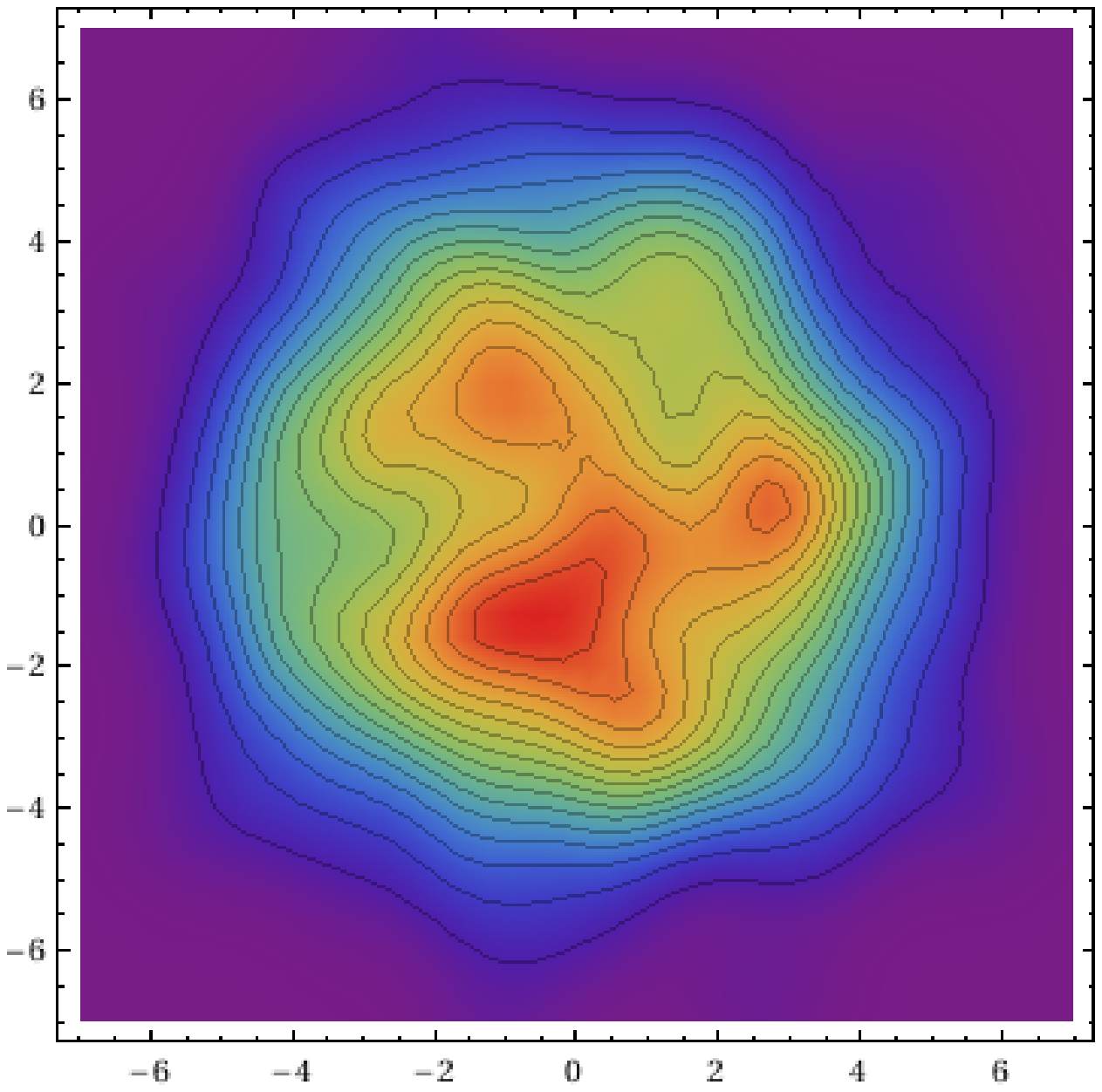}}
	\put(45,46){\includegraphics[width=4.2cm]{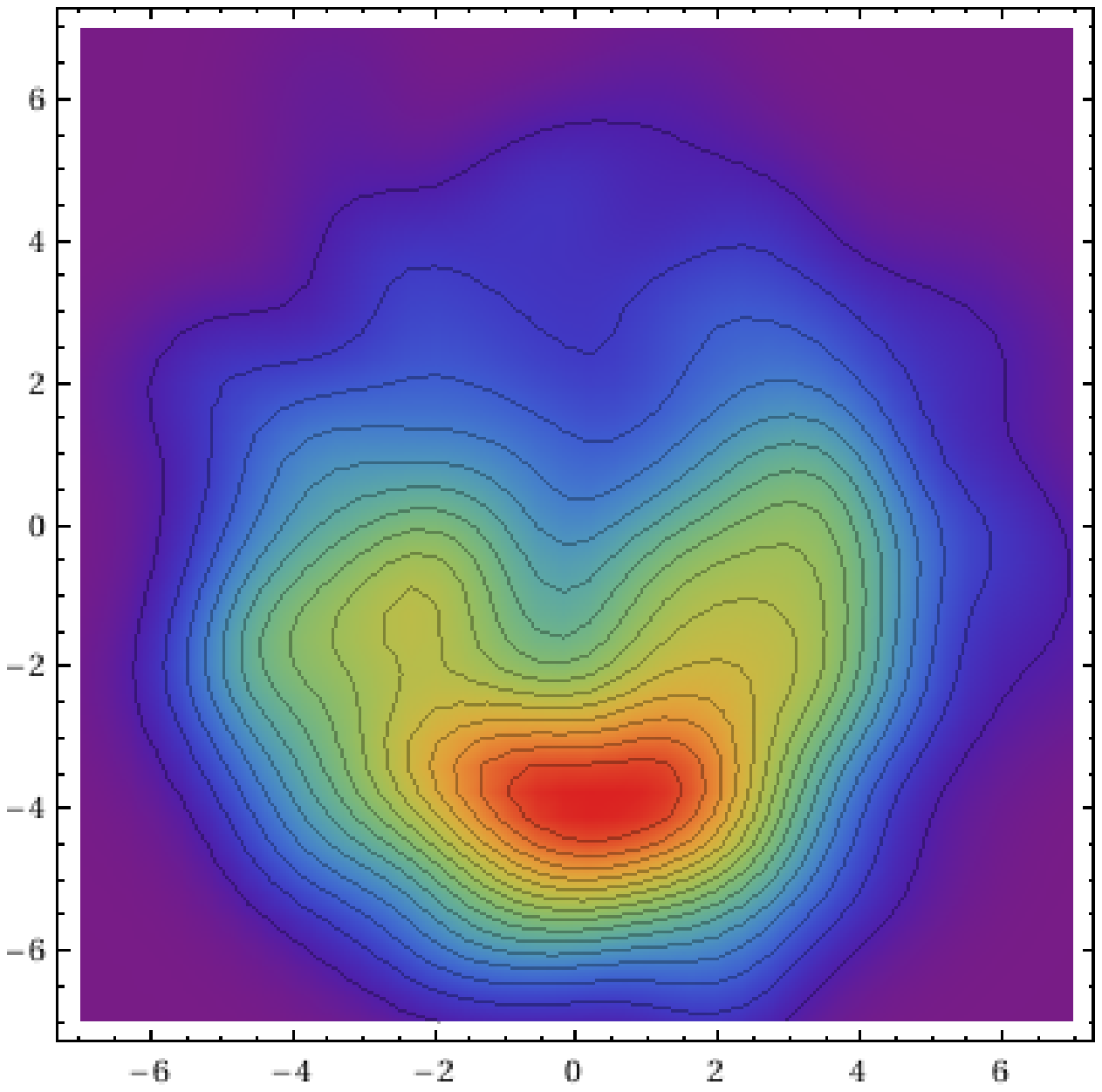}}

	\put(2,2){\includegraphics[width=4.2cm]{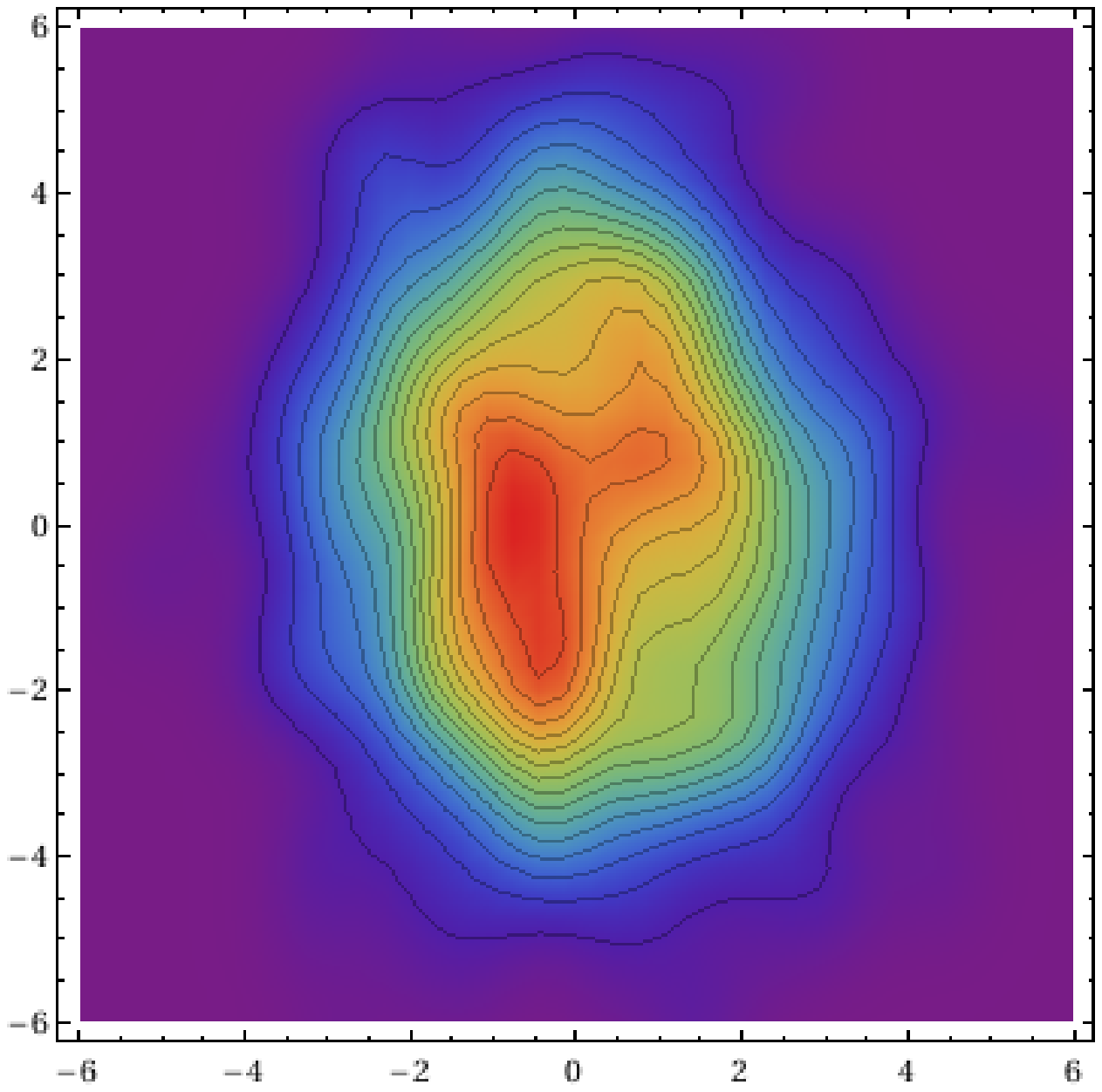}}
	\put(45,2){\includegraphics[width=4.2cm]{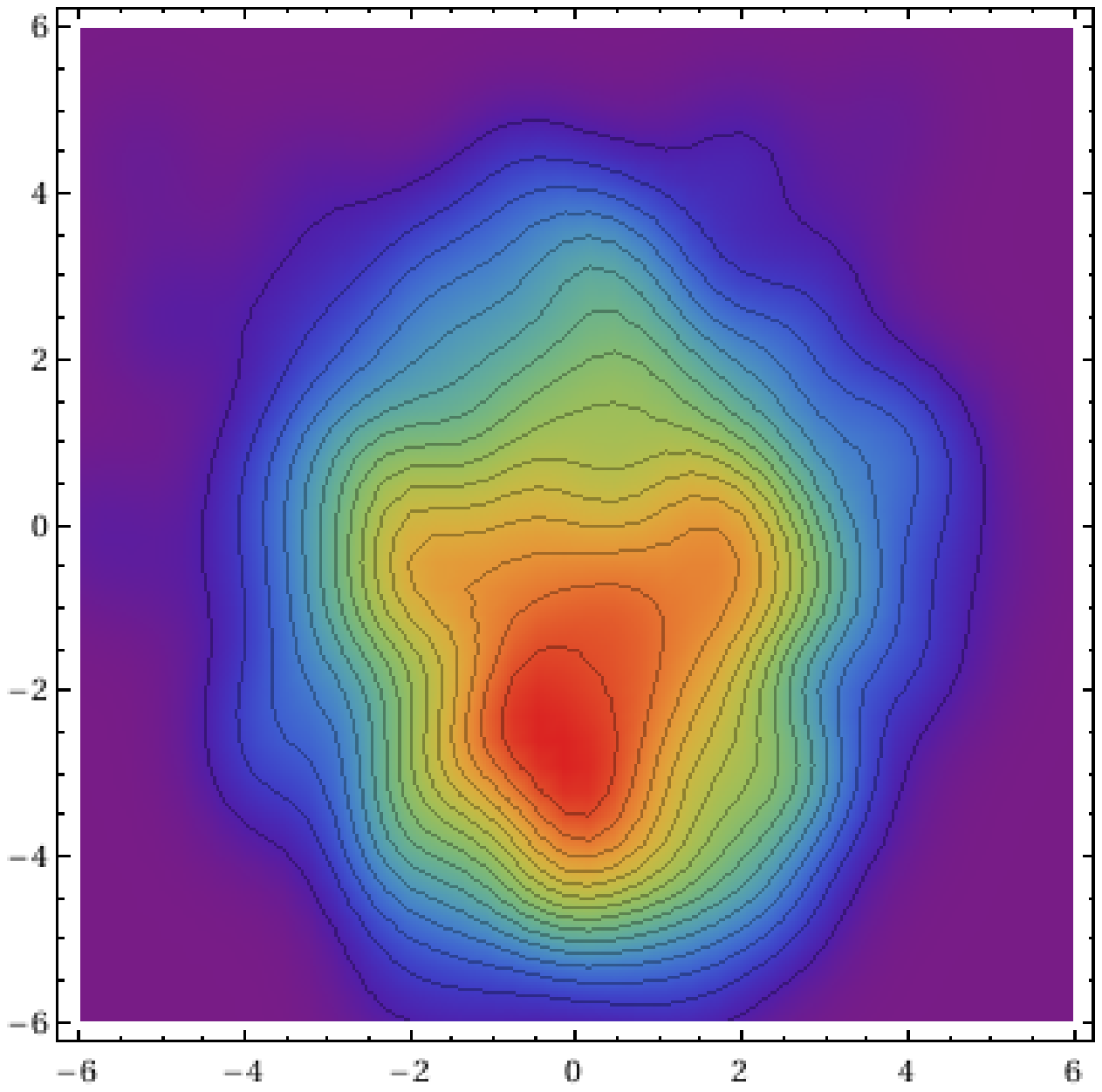}}

	\put(-1,43){\rotatebox{90}{y(fm)}}

	\put(37,82){\textcolor{black}{(a)}}
	\put(80,82){\textcolor{black}{(b)}}

	\put(37,38){\textcolor{black}{(c)}}
	\put(80,38){\textcolor{black}{(d)}}

	\end{picture}

	\caption{(Color online) The density distribution of the initial photon-jet production points $(x_{\rm ini}, y_{\rm ini})$ when the triggered photons are taken along out-of-plane directions ($|\phi_\gamma - \pi/2| < \pi/12$) for Pb+Pb collisions at the LHC. Jets with different final $x_T$ values are compared: $x_T = [0.5, 0.6]$ for (a) and (c); $x_T = [0.9, 1]$ for (b) and (d). Two different centralities are compared: $0-10\%$ for (a) and (b); $20-30\%$ for (c) and (d). The jet size is $R=0.3$.
	} \label{gammajet_xTlargesmall}
\end{figure}

\begin{figure}[tb]
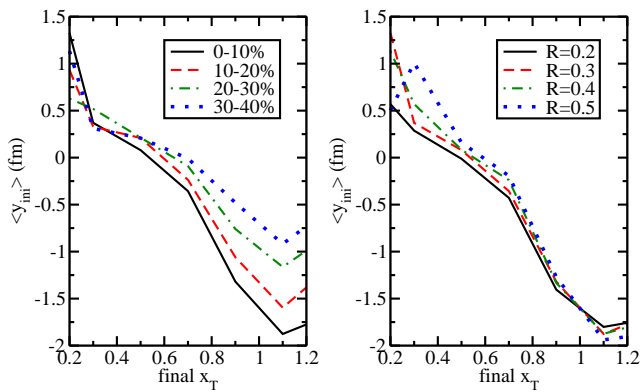

\includegraphics[width=0.49\linewidth]{6a_gammajet_tomography2.eps}
\includegraphics[width=0.49\linewidth]{6b_gammajet_tomographyR2.eps}
\caption{(Color online) The average values of the jet initial production points $\langle y_{\rm ini}\rangle$ when the triggered photons are taken along out-of-plane directions ($|\phi_\gamma - \pi/2| < \pi/12$) for Pb+Pb collisions at the LHC. The left show the results for different centralities with jet size $R=0.3$. The right shows the results for most central ($0-10\%$) collisions with different jet sizes.
} \label{gammajet_tomography}
\end{figure}

We have presented above the medium modification of the photon-triggered jet yield, the photon-jet momentum imbalance, and their azimuthal angle distribution for Pb+Pb collisions at the LHC.
We may further study photon-tagged jets by classifying the events according to their final $x_T$ values.
In such way, one may use photon-tagged jets to probe different regions of the hot and dense QGP created in the collisions.
This can be clearly seen in Fig. \ref{gammajet_xTlargesmall}, where we show the density distribution of initial photon-jet production points $(x_{\rm ini}, y_{\rm ini})$ in the transverse plane.
Here we take those events with triggered photons that propagate along out-of-plane directions ($|\phi_\gamma - \pi/2| < \pi/12$).
We compare the initial photon-jet production point distribution for different $x_T$ values: $x_T = [0.5, 0.6]$ for (a) and (c); $x_T = [0.9, 1]$ for (b) and (d). Results for two different centralities are also shown for comparison: $0-10\%$ for (a) and (b); $20-30\%$ for (c) and (d).
If one takes photon-jet events for all values of $x_T$ and propagating directions, the density distribution of initial photon-jet production points would follow the binary collision distribution (if no kinematic cuts are applied after medium evolution).
From the plots we may see that jets with larger values of final $x_T$ typically have traversed shorter medium length than smaller $x_T$ jets.

The above effect can be made more quantitative. In Fig. \ref{gammajet_tomography}, we show the average for the photon-jet initial production points $\langle y_{\rm ini}\rangle$ if one looks at the triggered photons that propagate along out-of-plane directions ($|\phi_\gamma - \pi/2| < \pi/12$).
One may see that on average jets with smaller $x_T$ values may have traversed about a few ${\rm fm}$ longer distance than larger $x_T$ jets.
Such medium length difference traversed by different $x_T$ jets is found larger in more central collisions (the left panel) and for larger jet cone sizes (the right panel).

\begin{figure}[tb]
\includegraphics[width=0.95\linewidth]{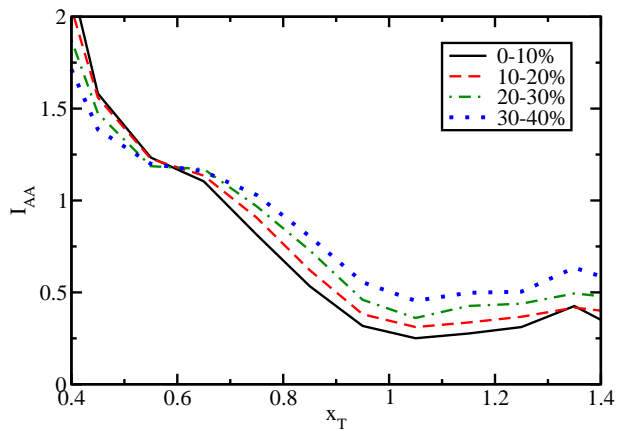}
\caption{(Color online) The nuclear modification factor $I_{AA}$ for photon-triggered jets as a function of $x_T$ for Pb+Pb collisions at the LHC in different centralities. The jet size is $R=0.3$.
} \label{gammajet_Iaa_vsxT}
\end{figure}

\begin{figure}[tb]
\includegraphics[width=0.95\linewidth]{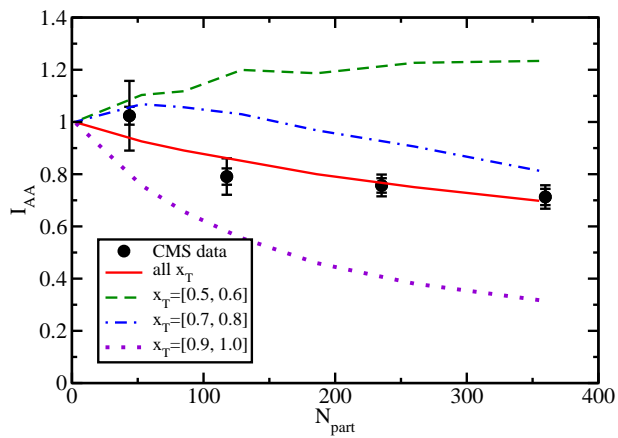}
\caption{(Color online) The nuclear modification factor $I_{AA}$ for photon-triggered jets as a function of centrality for Pb+Pb collisions at the LHC. Results for different $x_T$ values are compared. The jet size is $R=0.3$.
} \label{gammajet_Iaa_xT_vsnpart}
\end{figure}

Since jets with different $x_T$ values have traversed different medium lengths and density profiles (and lost different amount of energy), they are expected to exhibit different medium modification patterns.
In Fig. \ref{gammajet_Iaa_vsxT}, the photon-triggered nuclear modification factor $I_{AA}$ is plotted as a function of the momentum fraction $x_T$ for Pb+Pb collisions at the LHC.
One may observe that the yield for the associated jets with large $x_T$ values is suppressed in Pb+Pb collisions due to the interaction with the QGP medium, while the yield for lower $x_T$ jets is enhanced due to the fact that the momentum fraction variable $x_T$ distribution shifts from higher $x_T$ to lower $x_T$ values.
Such a medium modification effect is found to increase from peripheral collisions to central collisions.

The centrality dependence can be more clearly seen in Fig. \ref{gammajet_Iaa_xT_vsnpart}, where we show the $x_T$-integrated photon-triggered jet $I_{AA}$ as a function of $N_{\rm part}$ for Pb+Pb collisions at the LHC.
Results for different values of $x_T$ bins are shown for comparison.
Due to the combinational effects of different traveling distances and the tagged jet momentum spectra, we observe larger suppression and stronger centrality dependence for larger values of $x_T$, and smaller suppression (or enhancement) for smaller $x_T$ jets.

\begin{figure}[tb]
\includegraphics[width=0.95\linewidth]{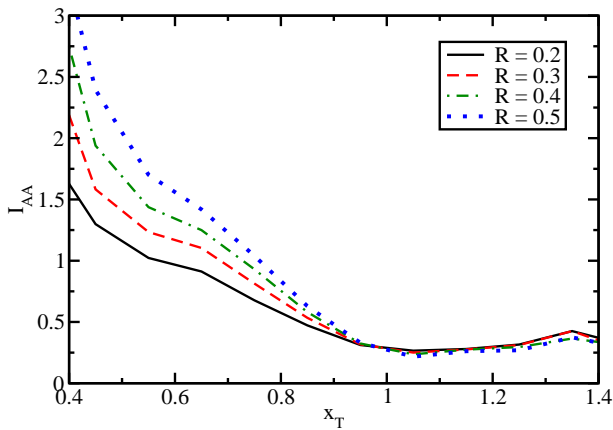}
\caption{(Color online) The nuclear modification factor $I_{AA}$ for photon-triggered jets as a function of $x_T$ for most central ($0-10\%$) Pb+Pb collisions at the LHC. Results for different jet sizes are compared.
} \label{gammajet_Iaa_vsxTR}
\end{figure}

\begin{figure}[tb]
\includegraphics[width=0.95\linewidth]{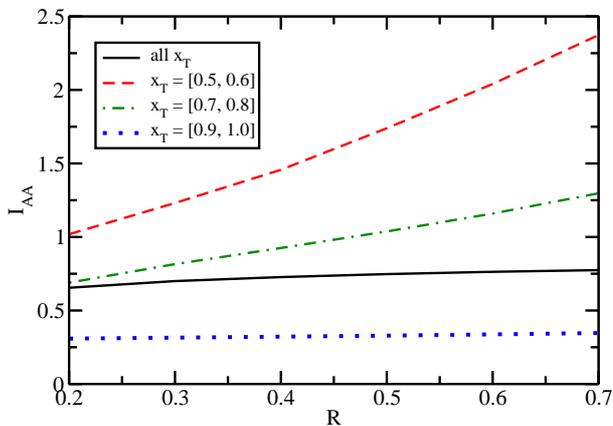}
\caption{(Color online) The jet size dependence of photon-triggered jet $I_{AA}$ for most central (0-10\%) Pb+Pb collisions at the LHC. Results for different $x_T$ values are compared.
} \label{gammajet_Iaa_vsR}
\end{figure}

The medium modification of photon-tagged jets depends on the cone sizes as well. In Fig. \ref{gammajet_Iaa_vsxTR}, the photon-tagged nuclear modification factor $I_{AA}$ is plotted as a function of $x_T$ for most central ($0-10\%$) Pb+Pb collisions at the LHC. Results for different jet sizes ($R=0.2, 0.3, 0.4, 0.5$) are shown for comparison.
We observe that the suppression of the associated jet yield at high $x_T$ is similar for different jet sizes, whereas the enhancement of small $x_T$ jet yield is larger when jet cone size is increased.
This indicates that the medium modification of smaller $x_T$ jets are more sensitive to the change of jet cone size due to traversing longer medium lengths.

We may see such cone size dependence due to medium path lengths in Fig. \ref{gammajet_Iaa_vsR}, where we show the $x_T$-integrated nuclear modification factor $I_{AA}$ as a function of jet cone size for most central (0-10\%) Pb+Pb collisions at the LHC. Results for different values of $x_T$ bins are shown for comparison.
The nuclear modification of larger $x_T$ jets show weak jet cone size dependence, while a stronger dependence is observed for smaller $x_T$ jets.

\begin{figure}[tb]
\includegraphics[width=0.95\linewidth]{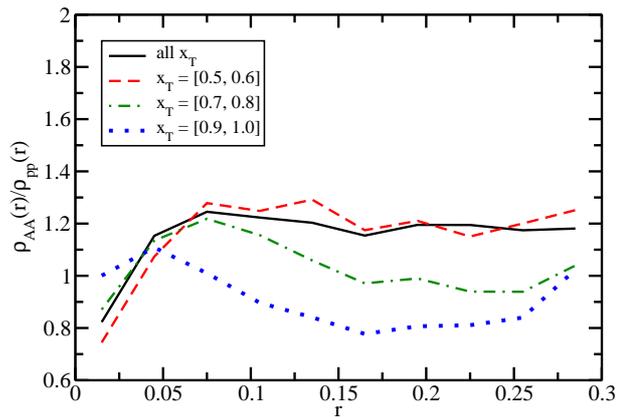}
\caption{(Color online) The nuclear modification of photon-triggered jet shape distribution $\rho(r)$ ($R=0.3$) for most central (0-10\%) Pb+Pb collisions at the LHC. Results for different $x_T$ values are compared.
} \label{gammajet_Dr_xT}
\end{figure}

It is also of great interest to study the nuclear modification of jet shape observables.  The jet shape $\rho(r)$ in the transverse direction is defined as
\begin{eqnarray}
\rho(r) = \frac{1}{p_{T,J}} \frac{\sum_{i} {p_{T,i}}(r_i \in [r - \Delta r/2, r + \Delta r /2])}{\Delta r},
	\end{eqnarray}
	where $p_{T, J}$ is the transverse momentum of the reconstructed jet.
	In the above equation, $p_{T,i}$ represents the transverse momentum of the individual partons within the jet, $r_i = \sqrt{(\phi_i - \phi_J)^2 + (\eta_i - \eta_J)^2}$, and $\Delta r$ is the bin size.
The jet shapes for single inclusive jets have been recently measured at the LHC by CMS Collaboration \cite{Chatrchyan:2013kwa}.
A redistribution of the jet energy inside the cone has been observed in Pb+Pb collisions as compared to p+p collisions: an excess of energy fraction at large $r$, a depletion at intermediate $r$ and little modification at very small $r$.

Here we study the jet shape observables for jets tagged by direct photons. 
In Fig. \ref{gammajet_Dr_xT}, we show the nuclear modification of jet shapes $\rho(r)$ of photon-tagged jets for different values of $x_T$.
While some broadening effect (i.e., an excess at large $r$) is observed for small $x_T$ jets, for the large $x_T$ jets, jet shape $\rho(r)$ with large $r$ values might be suppressed.
This may be understood from the different lengths traversed by different $x_T$ jets. As jets propagate through the medium, the gluon radiation (vacuum and medium-induced) at larger angles (larger $r$) interact with medium longer due to smaller formation times, while at later time, jet broadening effect takes over, leading to the enhancement at larger $r$ values for smaller $x_T$ jets.
These results indicate that jet shapes are sensitive to many details of full jets and their interaction with medium constituents.
It is interesting to note that Refs. \cite{MehtarTani:2010ma,MehtarTani:2011tz,CasalderreySolana:2012ef} have argued that the jet is mostly a coherent core, and thus loses energy in medium like a single parton without modifying jet inner structure.
We note that the hydrodynamic response of the medium to jet transport and the redistribution of lost energy may influence the final observed jet structure.
The inclusion of this contribution would be essential before performing quantitative comparison with the jet structure measurements. This will be pursued in future effort.

\section{Dependence on jet transport coefficients}

Jet transport coefficients encode information about the nature and properties of hot and dense matter probed by high energy jets.
In the previous section, we related jet transport coefficients to the medium temperature according to $\hat{q} \propto T^3$ and we assumed the relative sizes of the longitudinal and transverse transport coefficients as $\hat{q}=2\hat{e}_2=4T\hat{e}$.
In general, the relative sizes of longitudinal and transverse transport coefficients may depend on detailed structure of the medium, such as the mass of medium constituents \cite{ColemanSmith:2012xb}.
Jet transport coefficients may depend on the properties of the probes as well, such as the energy/momentum of the propagating jet showers.
In this section, we investigate how the medium modification of photon-tagged jets depends on the relative sizes and different parameterizations of transverse and longitudinal jet transport coefficients.

\begin{figure}[tb]
\includegraphics[width=0.95\linewidth]{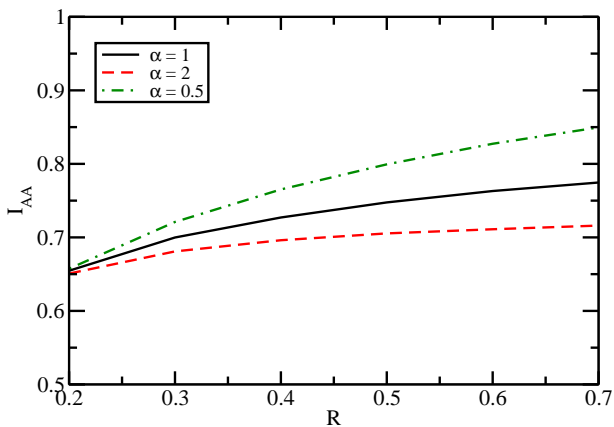}
\caption{(Color online) The jet size dependence of photon-triggered jet $I_{AA}$ for most central (0-10\%) Pb+Pb collisions at the LHC. Results for various values of $\alpha \equiv 4T\hat{e}/\hat{q}$ are compared.
} \label{gammajet_Iaa_vsRehat}
\end{figure}

\begin{figure}[tb]
\includegraphics[width=0.95\linewidth]{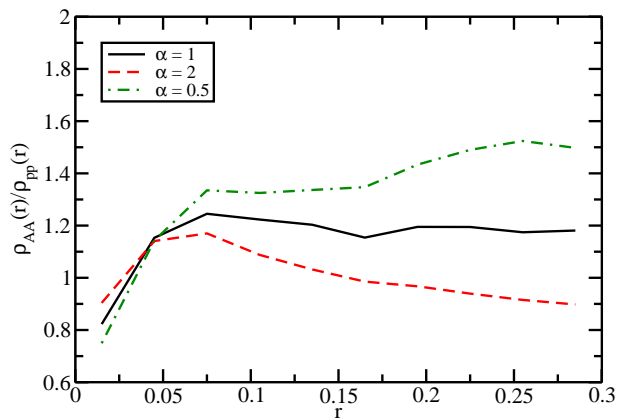}
\caption{(Color online) The nuclear modification of photon-triggered jet shape distribution $\rho(r)$ ($R=0.3$) for most central (0-10\%) Pb+Pb collisions at the LHC.
Results for various values of $\alpha \equiv 4T\hat{e}/\hat{q}$ are compared.
} \label{gammajet_Dr_ehat}
\end{figure}

We first look at the dependence of photon-tagged jet nuclear modification of the relative sizes between transverse and longitudinal jet transport coefficients.
Here the transport coefficients are still related to the medium temperature according to $\hat{q} \propto T^3$.
In Fig. \ref{gammajet_Iaa_vsRehat}, we show the $x_T$-integrated photon-tagged jet $I_{AA}$ as a function of the jet cone size for most central (0-10\%) Pb+Pb collisions at the LHC. The results for different values of $\alpha \equiv 4T\hat{e}/\hat{q}$: $\alpha = 2, 1, 0.5$ are shown for comparison.
To see the pure effect on jet cone size dependence more clearly, we have refitted photon-tagged jet $I_{AA}$ to the same values for $R=0.2$ (the smallest jet size that experiments use) when changing the relative sizes of $\hat{e}$ and $\hat{q}$.
With the increase of the relative size of $\hat{q}$, a stronger jet cone size dependence is observed for the nuclear modification $I_{AA}$ of the photon-tagged jet yield.
This is due to the fact that the relative contribution from medium-induced radiation and jet broadening effect become larger when increasing the relative size of $\hat{q}$ (the decrease of $\alpha = 4T\hat{e}/\hat{q}$).

In Fig. \ref{gammajet_Dr_ehat}, we show the nuclear modification of the jet shape observable $\rho(r)$ for photon-triggered jets as a function of jet cone size for most central (0-10\%) Pb+Pb collisions at the LHC. A larger broadening effect is seen as we increase the relative size of $\hat{q}$. We also observe some suppression at larger values of $r$ when $\hat{q}$ becomes small ($\hat{e}$ becomes large). For such case, the energy loss of jet showers is more dominated by the collisional energy loss; the radiated gluons at larger angles interact with medium earlier due to their smaller formation times.

\begin{figure}[tb]
\includegraphics[width=0.95\linewidth]{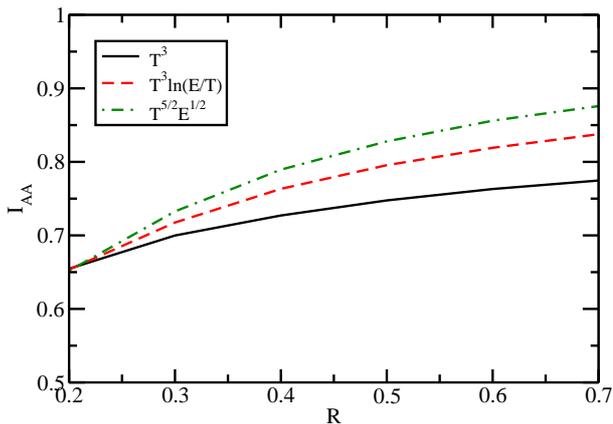}
\caption{(Color online) The jet size dependence of photon-triggered jet $I_{AA}$ for most central (0-10\%) Pb+Pb collisions at the LHC.
Results for different parameterizations of $\hat{q}$ are compared.
} \label{gammajet_Iaa_vsRqhat}
\end{figure}

\begin{figure}[tb]
\includegraphics[width=0.95\linewidth]{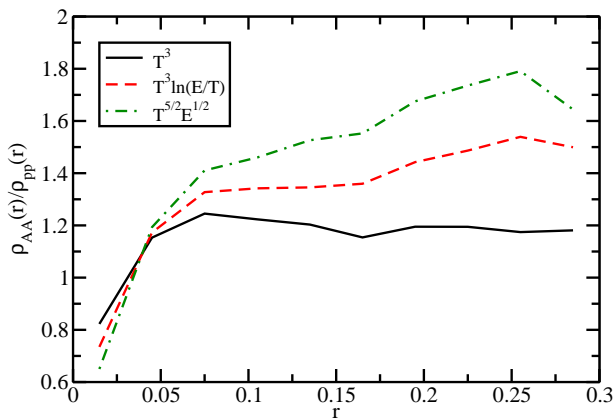}
\caption{(Color online) The nuclear modification of photon-triggered jet shape distribution $\rho(r)$ ($R=0.3$) for most central (0-10\%) Pb+Pb collisions at the LHC.
Results for different parameterizations of $\hat{q}$ are compared.
} \label{gammajet_Dr_qhat}
\end{figure}

We now investigate the dependence of photon-tagged jet modification on the different parameterizations of jet transport coefficients.
For such purpose, we fix the relative sizes between longitudinal and transverse transport coefficients as $\hat{q}=4T\hat{e}$.
We compare three different parameterizations: $\hat{q} \propto T^3$, $T^3 \ln(E/T)$ and $T^{5/2}E^{1/2}$ to study the influence of energy dependence of jet transport coefficients.
The energy dependence of the form $T^3 \ln(E/T)$ originates from the leading-log Hard-Thermal-Loop results \cite{Qin:2007rn}, and the use of power dependence on energy is motivated from a lattice calculation of jet quenching parameter \cite{Majumder:2012sh}.

In Fig. \ref{gammajet_Iaa_vsRqhat}, we show the $x_T$-integrated photon-triggered jet $I_{AA}$ as a function of jet cone size for most central (0-10\%) Pb+Pb collisions at the LHC.
Again, we have refitted the value of $I_{AA}$ to the same values for $R=0.2$ when different parameterizations are used.
One can see that the power law ($E^{1/2}$) parametrization gives the strongest jet cone size dependence for photon-tagged jet $I_{AA}$.
This may be understood from the fact that the radiated gluons with smaller energies are more difficult to interact with the medium constituents than much higher energy leading parton, thus the energy loss of jet shower is more dominated by the medium-induced radiation as compared to the other two scenarios.
In Fig. \ref{gammajet_Dr_qhat}, we show the nuclear modification of jet shape observable $\rho(r)$ for photon-triggered jets as a function of jet cone size for most central (0-10\%) Pb+Pb collisions at the LHC. Again a larger jet broadening effect is observed as we increase the energy dependence for jet transport coefficients.

\section{Summary}

We have studied the medium modification of jets correlated with high $p_T$ photons in Pb+Pb collisions at the LHC.
A transport and perturbative QCD hybrid model is developed for simulating jet shower evolution and modification in hot and dense QCD medium with the inclusion of the contributions from both induced gluon radiation and elastic collisions.
The effect of elastic collisions experienced by jet shower partons is encoded by a few transport coefficients, while medium-induced gluon radiation is simulated with the use of the single gluon emission spectrum obtained from higher twist jet energy loss calculations.

With our event-by-event parton shower simulation, we have presented the numerical results for the nuclear modification of the photon-tagged jet production yield, the photon-jet momentum imbalance and their azimuthal angle distribution.
We have studied photon-tagged jets with different values of the momentum fraction $x_T$, and they exhibit different medium modification patterns due to traversing different medium lengths and density profiles.
This makes photon-triggered jets very useful for tomographic study of jet modification in dense matter created in relativistic nuclear collisions.

We have further investigated the influence of transverse and longitudinal jet transport coefficients on the modification of photon-tagged jets and found that both the relative sizes and different parametrization forms of $\hat{e}$ and $\hat{q}$ may affect the nuclear modification patterns of photon-tagged jets, especially the jet cone size dependence and jet shape observables.
Such results indicate that the study of cone size dependence and jet shape observables may provide strong constraints on the quantitative extraction of transverse and longitudinal jet transport coefficients. The determination of these coefficients will provide insights into the internal structure and various transport properties of the hot and dense medium produced in relativistic heavy-ion collisions.

The above analysis may be applied for studying nuclear modification of high $p_T$ dijets and single inclusive jets.
The systematic comparison to various experimental observables such as those done in Ref. \cite{Renk:2011aa} will provide more physics on jet modification.
The inclusion of a hadronization process for final partonic jet showers will enable us to directly compare to the experimental measurements of the longitudinal momentum (fraction) distribution of hadron fragments of jets.
The energy and momentum deposition profiles obtained from our simulation of in-medium jet shower evolution may be applied in hydrodynamics simulation to investigate the response of the medium to jet transport. 
Incorporating both jet energy loss and medium response components is essential to perform more quantitative comparison with final observed jet structure and correlations.
These studies will be very helpful for more complete understanding of jet-medium interaction and are left for future effort.

\begin{acknowledgements}

We would like to thank the Ohio State University group (U. Heinz, Z. Qiu, C. Shen and H. Song) for providing the corresponding initialization and hydrodynamic evolution codes.
We thank S. Cao, B. M\"uller and X.-N. Wang for helpful discussions. We thank C. Coleman-Smith and Majumder for careful reading of the manuscript and for discussions.
This work was supported in part by the Natural Science Foundation of China under grant no 11375072 and the U. S. Department of Energy under grants DE-FG02-05ER41367.

\end{acknowledgements}

\bibliographystyle{spphys}       
\bibliography{GYQ_refs}

%
%

\end{document}